\documentclass[aip, apl, superscriptaddress, reprint]{revtex4-1}

\usepackage{amsmath}
\usepackage{graphicx}

\begin{document}

\title{qPlus Magnetic Force Microscopy in Frequency-Modulation Mode with milli-Hertz Resolution}

\date{March 2011}

\author{Maximilian Schneiderbauer}
\author{Franz J. Giessibl}
\affiliation{Institute of Experimental and Applied Physics, University of Regensburg, 93040 Regensburg, Germany}

\begin{abstract}
Magnetic force microscopy (MFM) allows one to image the domain structure of ferromagnetic samples by probing the dipole forces between a magnetic probe tip and a magnetic sample. The magnetic domain structure of the sample depends on the atomic arrangement of individual electron spins. It is desirable to be able to image both individual atoms and domain structures with a single probe. However, the force gradients of the interactions responsible for atomic contrast and those causing domain contrast are orders of magnitude apart - ranging from up to 100\,Nm\textsuperscript{-1} for atomic interactions down to 0.0001\,Nm\textsuperscript{-1} for magnetic dipole interactions. Here, we show that this gap can be bridged with a qPlus sensor, with a stiffness of 1\,800\,Nm\textsuperscript{-1} (optimized for atomic interaction), that is sensitive enough to measure milli-Hertz frequency contrast caused by magnetic dipole-dipole interactions. Thus we have succeeded to establish a sensing technique that performs Scanning Tunneling Microscopy, Atomic Force Microscopy and MFM with a single probe.
\end{abstract}

\maketitle

\begin{figure}
  \centering
  \includegraphics[width=\columnwidth]{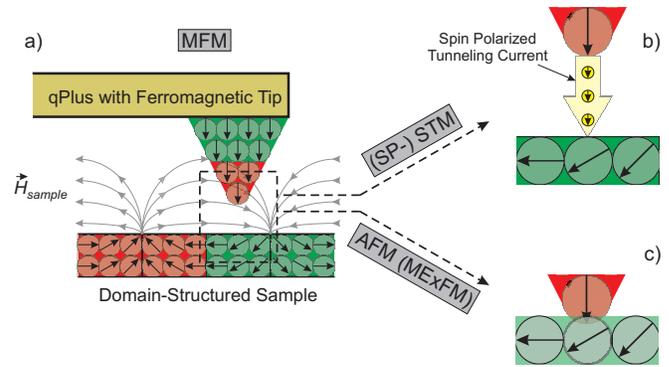}
  \caption{a) MFM probes the force between the magnetic dipole moment of a probe tip and the magnetic stray field of a sample. With a qPlus sensor, the same probe can be used to perform b) (SP-) STM and c) AFM (MExFM) experiments.}
  \label{big3}
\end{figure}

Solid state magnetism plays an important role in todays life, mainly for data storage technology. Ferromagnetism has its origin in the parallel alignment of atomic magnetic dipole moments, which is primarily given by the electron spin of an atom, and is therefore a collective phenomenon of atoms. Whereas the classical magnetic dipole interaction is far too weak to mediate this, it can be explained by the quantum mechanical exchange interaction. Regions of aligned spins, called domains, are used for example to store bits on hard disks. Such ferromagnetic domains have much larger magnetic dipole moments, as many atoms contribute to the resulting moment.\\
To probe magnetic structures on the atomic as well as on the domain size level in real space, variations of Scanning Tunneling Microscopy (STM) \cite{Binnig1982} and  Atomic Force Microscopy (AFM) \cite{Binnig1986} are used. To explore spin structures on conductive samples, the Spin Polarized-STM (SP-STM) \cite{Wiesendanger1990, Wiesendanger2009} is a powerful tool. The SP-STM measures the spin-dependent conductivity between a spin-polarized tip and the spin dependent local density of states of the sample [Fig. \ref{big3}.b)]. STM is unable to probe insulating surfaces but AFM can be used; the antiferromagnetic surface structure of NiO (001) was recently imaged by Magnetic Exchange Force Microscopy (MExFM) \cite{Kaiser2007}. In MExFM the magnetic exchange force between a tip atom with fixed spin orientation and a sample atom is measured [Fig. \ref{big3}.c)].\\
Imaging magnetic domains by Magnetic Force Microscopy (MFM) \cite{Martin1987b, Saenz1987} is nowadays well established. MFM images the magnetic dipole interaction of a ferromagnetic tip and a domain structured sample [Fig. \ref{big3}.a)]. Typically, magnetically coated silicon cantilevers are used. These cantilevers are produced in large quantity by micro-fabrication techniques. Typical probe features are spring constants in the order of 10\,Nm\textsuperscript{-1} and resonance frequencies of about 100\,kHz. Another type of force sensors is made of quartz (SiO$_2$) tuning forks. The qPlus sensor \cite{Giessibl2000} is based on a quartz tuning fork, where one prong is attached to a carrier substrate. The qPlus's large spring constant of $k=1\,800$\,Nm\textsuperscript{-1} allows to overcome the snap-to-contact-problem in small amplitude operation \cite{Giessibl2000c}. In that, the qPlus setup is customized for combined STM/AFM measurements with atomic resolution \cite{G2002}.
However in standard MFM experiments, this large $k$, in combination with the resonance frequency $f_0 = 31\,000$\,Hz, leads to very small frequency shifts [Eq. (\ref{gradientapprox})].\\ 
The qPlus sensor has not yet proven its ability to detect the weak long-range magnetostatic interaction. In this article we show that the qPlus sensor is also capable of MFM experiments. We show imaging contrast of some milli-Hertz in the large amplitude regime, which is typically used for MFM. Therefore, we achieved a setup that is able to record a wide range of scanning probe imaging signals; starting from domain resolving MFM experiments, culminating in atomically resolved STM and AFM experiments [Fig. \ref{big3}].\\
In frequency modulation AFM (FM-AFM) the measured frequency shift $\Delta f$ is proportional to an averaged force gradient $\langle k_{ts} \rangle$ with $k_{ts}=-\partial F_{ts} / \partial z$; $F_{ts}$ is the force acting between tip and sample within one oscillation period; $z$-direction is perpendicular to the sample surface. Within the gradient approximation, $\Delta f$ is given by:
\begin{equation}
\Delta f = \frac{f_0}{2 k} \langle k_{ts} \rangle
\label{gradientapprox}
\end{equation}
The frequency noise $\delta(\Delta f)$ in FM-AFM setups determines the minimum detectable averaged force gradient $\langle  k_{ts} \rangle_{min}$, where $\delta(\Delta f)$ is a sum of three uncorrelated noise sources \cite{Albrecht1991, Kobayashi2009}.
Furthermore the frequency noise is inversely proportional to the force sensor's oscillation amplitude $A$. Thus, we can reduce frequency noise by using large amplitudes and therefore minimize the $\langle  k_{ts} \rangle_{min}$. Moreover, one gets the best signal-to-noise ratio using amplitudes in the order of the decay length of the interaction to be measured \cite{Giessibl1999}. The magnetic dipole force in MFM has a decay length in the range of domain sizes, around 100\,nm.\\
Typical values in our ambient qPlus setup are $f_0 = 31\,000$\,Hz, $k=1\,800$\,Nm\textsuperscript{-1}, quality factor $Q=2\,000$, measurement bandwidth $B = 120$\,Hz and deflection noise density $n_q=100\,$fm/$\sqrt{\textnormal{Hz}}$. To maximize the sensitivity and the signal-to-noise ratio we chose an amplitude of $A=100$\,nm and obtain a frequency noise of $\delta(\Delta f) = 1.14$\,mHz. 
From Equation (\ref{gradientapprox}) we can calculate the minimum detectable force gradient $\langle k_{ts} \rangle_{min} = 1.32\times 10^{-4}$\,Nm\textsuperscript{-1}. Commercial MFM cantilevers, like the PPP-MFMR from Nanosensors, in a standard setup are sensitive to force gradients down to $\langle k_{ts} \rangle_{min} = 4.87\times 10^{-7}$\,Nm\textsuperscript{-1}.\\

All experiments presented here were done in ambient conditions. For vibration isolation the microscope is mounted on a mechanical double damping stage \cite{Park1987}. We used the Nanonis SPM \cite{Nanonis} control electronics and the Multipass configuration to perform Lift Mode experiments for MFM.
The Lift Mode is a two pass technique which enables a separation of topographic and, here, magnetic signals. In the first pass the surface is scanned in non-contact AFM to obtain the topography of the surface. Within the second pass, the topographic lines are used to track the probe in an elevated tip-sample distance over the surface. Thus, the van der Waals force is kept constant and any force change is caused by the long-range magnetostatic interaction.
For FM detection we utilized a homebuilt amplitude controller and for frequency demodulation the Nanosurf easy PLL was used.
As a reference sample we used a 41\,GB hard disk from MAXTOR with a bit-density of approximately 2\,Gbit/in\textsuperscript{2}.\\
The magnetostatic force is a function of the tip's magnetic moment and the gradient of the surface's magnetic stray-field \cite{Hartmann1999}:
\begin{equation}
\vec{F}_{mag} = \mu_0 (\vec{m}_{tip} \cdot \nabla ) \vec{H}_{sample}
\label{Fmag}
\end{equation} 
Here $\vec{m}_{tip}$ is the probe's effective dipole moment and $\vec{H}_{sample}$ is the sample's magnetic stray field. As $\vec{H}_{sample}$ primarily varies in $z$-direction, perpendicular to the sample surface, the main contribution of $F_{mag}$ is given by the partial derivative in $z$-direction. By using the same sample one can therefore vary the interaction strength by the tip's magnetic moment and the Lift Mode height.\\

\begin{figure*}
  \centering
  \includegraphics[width=14cm]{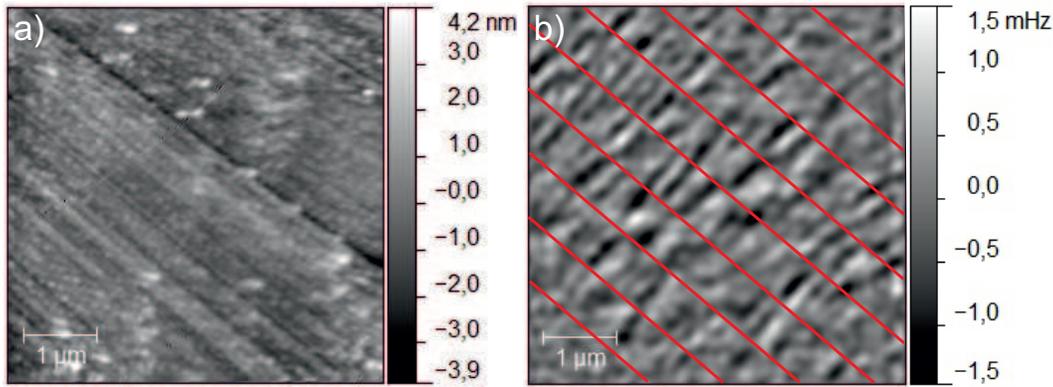}
  \caption{Lift Mode FM-MFM image using a qPlus sensor with an etched iron tip attached to it. Flattened raw data with imaging parameters $f_0 = 31\,679$\,Hz, $k=1\,800$\,Nm\textsuperscript{-1}, $Q=1\,337$, $A= 100$\,nm and lift height 50\,nm. a) Topography and b) Lift Mode frequency shift (low pass filtered); red lines as a guide for your eyes to the bit-tracks.}
  \label{dfiron}
\end{figure*}

\begin{figure*}
  \centering
  \includegraphics[width=14cm]{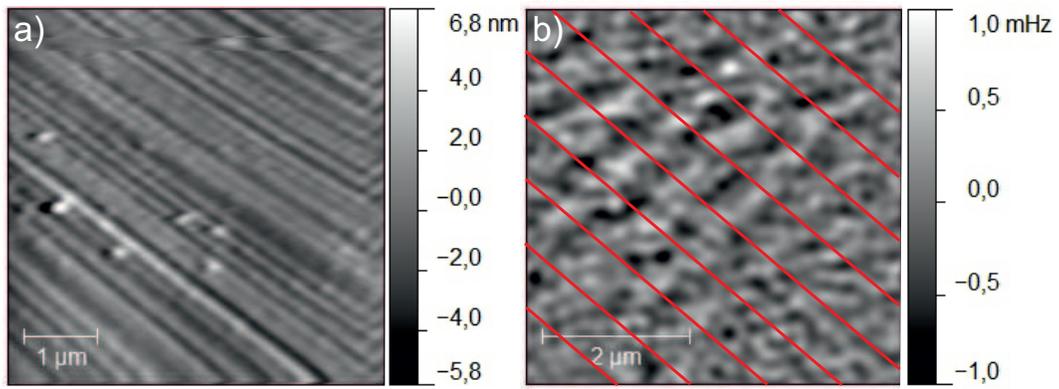}
  \caption{Lift Mode FM-MFM image using a qPlus sensor with a commercial cobalt-coated MFM cantilever tip attached to it. Flattened raw data with imaging parameters $f_0 = 31\,453$\,Hz, $k=1\,800$\,Nm\textsuperscript{-1}, $Q=1\,625$, $A= 100$\,nm and lift height 50\,nm. a) Topography and b) Lift Mode frequency shift (low pass filtered); red lines as a guide for your eyes to the bit-tracks.}
  \label{dfcantilever}
\end{figure*}

In a first attempt we used an electro-chemical etched bulk-iron tip and magnetized it for scanning with a strong permanent magnet. With this tip and a lift height of 50\,nm we have imaged the bit structure of the hard disk sample. Results were confirmed by scanning the same sample with a commercial silicon MFM cantilever setup (Nanosurf Flex AFM). Furthermore we deduced the correct bit density from the recorded magnetic structure in Fig. \ref{dfiron}.b) of $\approx$ 1.9\,Gbit/in\textsuperscript{2}. The topographic image shows the typical surface texture of a hard disk [Fig. \ref{dfiron}.a)]. The sizeable drift in both images is due to long measuring times, which were necessary to reduce the noise by reducing the bandwidth. According to the bit density (approximate bit size of 200\,nm by 600\,nm) and scan speed of 2,5\,$\mu$m/s, we expect an effective signal bandwidth of $\approx$ 10\,Hz in the frequency shift data set. Therefore we applied a 2D FFT filter with cut-off frequency of 20\,Hz to get rid of high frequency noise and thus increasing the signal-to-noise ratio. In Fig. \ref{dfiron}.b) the flattened and filtered frequency shift data gathered in Lift Mode show a image contrast of $\pm$1.5\,mHz; red lines mark the borders of single bit-tracks.\\
As large magnetic moments of the probe's tip can influence and even destroy the magnetic structure of the observed sample, a small magnetic tip moment is desirable. However, tips with a small magnetic moment reduce the interaction energy [Eq. (\ref{Fmag})] and thus the signal strength, bringing the signal close to its noise floor. Here a trade-off has to be made between increased sensitivity due to decreased measurement bandwidth and large thermal drift, at room temperature, due to long acquisition times. Therefore it is useful to apply a low pass filter in the spatial frequency regime of the acquired data set.\\
To benchmark our setup, we reduced the magnetic moment of the tip by gluing a commercial MFM cantilever tip (NanoWorld Pointprobe MFMR-10 Cobalt coated) onto a qPlus sensor. This has been done before in tuning fork setups at room temperature ultra-high-vacuum systems \cite{Rozhok2002} and low temperature systems \cite{Seo2005, Callaghan2005, Kim2006a}. First pass topography data set shows the expected surface structure [Fig. \ref{dfcantilever}.a)]. The scan speed had to be set to relative slow values, allowing for a small bandwidth, but leading to sizeable drift, as seen in Fig. \ref{dfcantilever}. As the spatial frequency of the sample did not change, we applied the same 2D FFT filter as in Fig. \ref{dfiron}.b) with bandwidth of 20\,Hz to the frequency shift data in Fig. \ref{dfcantilever}.b). This flattened and filtered Lift Mode data then revealed an image contrast of approximately $\pm$1.0\,mHz [Fig. \ref{dfcantilever}.b)]; borders of single bit-tracks are marked with red lines.\\

The spatial resolution and the signal-to-noise ratio gathered with etched iron tips are better than those taken with the MFM cantilever tip.\\

In summary, we have obtained an important benchmark: a probe that is optimized for high resolution AFM and STM can at the same time measure the tiny force gradients acting in MFM. We note that the experiments were performed in ambient conditions and we expect that future experiments in ultra-high-vacuum and low temperatures will enable us to measure even smaller force gradients.\\

We acknowledge the Sonderforschungsbereich 689 for financial support.

\end{document}